\documentstyle[epsfig]{l-aa}
\input{psfig}

\begin{document}

\thesaurus{11.02.1;11.02.2:BL~Lacertae;11.02.2:1ES 2344+51.4;13.07.2}

\title{HEGRA search for TeV emission from BL Lac objects}

\author{F.A. Aharonian\inst{1},
A.G. Akhperjanian\inst{7},
J.A.~Barrio\inst{2,3},
K.~Bernl\"ohr\inst{1,}$^*$,
H. Bojahr\inst{6},
I. Calle\inst{3},
J.L. Contreras\inst{3},
J. Cortina\inst{3},
A. Daum\inst{1},
T. Deckers\inst{5},
S. Denninghoff\inst{2},
V. Fonseca\inst{3},
J.C. Gonzalez\inst{3},
G. Heinzelmann\inst{4},
M. Hemberger\inst{1},
G. Hermann\inst{1,}$^\dag$,
M. He{\ss}\inst{1},
A. Heusler\inst{1},
W. Hofmann\inst{1},
H. Hohl\inst{6},
D. Horns\inst{4},
A. Ibarra\inst{3},
R. Kankanyan\inst{1,7},
M. Kestel\inst{2},
J. Kettler\inst{1},
C. K\"ohler\inst{1},
A. Konopelko\inst{1,}$^\S$,
H. Kornmeyer\inst{2},
D. Kranich\inst{2},
H. Krawczynski\inst{1,4},
H. Lampeitl\inst{1},
A. Lindner\inst{4},
E. Lorenz\inst{2},
N. Magnussen\inst{6},
O. Mang\inst{5},
H. Meyer\inst{6},
R. Mirzoyan\inst{2},
A. Moralejo\inst{3},
L. Padilla\inst{3},
M. Panter\inst{1},
D. Petry\inst{2,6,}$^\ddag$,
R. Plaga\inst{2},
A. Plyasheshnikov\inst{1,}$^\S$,
J. Prahl\inst{4},
G. P\"uhlhofer\inst{1},
G. Rauterberg\inst{5},
C. Renault\inst{1,}$^\#$,
W. Rhode\inst{6},
A. R\"ohring\inst{4},
V. Sahakian\inst{7},
M. Samorski\inst{5},
D. Schmele\inst{4},
M. Schilling\inst{5},
F. Schr\"oder\inst{6},
W. Stamm\inst{5},
H.J. V\"olk\inst{1},
B. Wiebel-Sooth\inst{6},
C. Wiedner\inst{1},
M. Willmer\inst{5},
W. Wittek\inst{2}}
\institute{Max Planck Institut f\"ur Kernphysik,
Postfach 103980, D-69029 Heidelberg, Germany \and
Max Planck Institut f\"ur Physik, F\"ohringer Ring
6, D-80805 M\"unchen, Germany \and
Universidad Complutense, Facultad de Ciencias
F\'{\i}sicas, Ciudad Universitaria, E-28040 Madrid, Spain \and
Universit\"at Hamburg, II. Institut fuer
Experimentalphysik, Luruper Chausse 149,
D-22761 Hamburg, Germany \and
Universit\"at Kiel, Institut f\"ur Experimentelle und Angewandte Physik,
Leibnizstra{\ss}e 17-19, D-24118 Kiel, Germany \and
Universit\"at Wuppertal, Fachbereich Physik,
Gau{\ss}str.20, D-42097 Wuppertal, Germany \and
Yerevan Physics Institute, Alikhanian Br. 2, 375036 Yerevan, Armenia\\
\hspace*{-4.04mm} $^*\,$ Now at Forschungszentrum Karlsruhe, P.O. Box
3640, D-76021 Karlsruhe, Germany\\
\hspace*{-4.04mm} $^\dag\,$ Now at Enrico Fermi Institute, University of
Chicago, 933 East 56th Street,
Chicago, IL 60637, U.S.A.\\
\hspace*{-4.04mm} $^\ddag\,$ Now at Universidad Aut\'{o}noma de Barcelona, 
Instituto de F\'{\i}sica d'Altes Energies, E-08193 Bellaterra, Spain\\
\hspace*{-4.04mm} $^\S\,$ \emph{On leave from:} Altai State
University, Dimitrov Street 66, 656099 Barnaul, Russia \\
\hspace*{-4.04mm} $^\#\,$ Now at LPNHE, Universit\'es Paris VI-VII, 4 place
Jussieu, F-75252 Paris Cedex 05, France
}

\offprints{C. Renault (rcecile@in2p3.fr)}

\date{Received 30 March 1999 / Accepted 12 November 1999} 

\maketitle
\markboth{Aharonian {\it et al.}: HEGRA search for TeV emission from BL Lac objects}{}

\begin{abstract} 

The HEGRA system of four Imaging Atmospheric
Cherenkov Telescopes (IACTs) has been used to extensively 
observe extragalactic objects. In this paper we describe the search
for TeV emission from nine very promising potential TeV sources, namely
eight ``high frequency'' BL Lac objects (HBLs), and the object ``BL Lacertae''  
itself. These objects were observed during 1997 and
1998 seasons, with total integration times ranging between one and fifteen hours.
 No evidence for emission was found from any of these objects
and the upper limits on the
integral energy flux above $\sim$750 GeV are on the level of a few times
\mbox{10$^{-12}$ erg~cm$^{-2}$~s$^{-1}$.} 
For the two objects BL Lacertae and 1ES 2344+51.4, we
discuss the astrophysical implications of the TeV flux upper limit, using
also information from the X-ray and $\gamma$-ray bands as measured with
the All Sky Monitor (ASM) of RXTE 
(1.3-12.0 keV) and with EGRET (30 MeV - 20 GeV).
\end{abstract}

\section{Introduction} 
%
The phenomenology of BL Lac objects has led to a classification of these 
sources into two sub-populations:
high-frequency BL Lac objects (HBLs) and low-frequency ones (LBLs)
(see {\it e.g.} Fossati {\it et al.} 1998). The HBL objects
have synchrotron and Compton peaks at relatively high energies (Ghisellini {\it et al.} 1998),
 and thus are good candidates for TeV emission.
With the HEGRA system of IACTs the significant detection
({\it e.g.} $>$ 3 $\sigma$) of a flux comparable to the flux from the Crab nebula takes less
than 15 minutes for sources close to zenith position.
 The BL Lac observations presented in this paper are thus
 sufficient to probe energy fluxes down to the level of \mbox{10$^{-12}$
erg~cm$^{-2}$~s$^{-1}$} ($\stackrel{>}{{}_{\sim}}750$ GeV).

The two BL Lac objects Mkn 501 and Mkn 421 are well established and well
studied TeV sources. Mkn 501 showed during 1997  spectacular outbursts
with diurnal flux levels reaching 10 Crab units (e.g., Aharonian {\it et al.}
1999a, Samuelson {\it et al.} 1998, Djannati {\it et al.} 1999) and with a spectrum extending up to at
least 16 TeV (Aharonian {\it et al.} 1999b). For a better
understanding of the intrinsic properties of this class of sources and for
determining the amount of intergalactic extinction of the TeV photons due
to pair production processes with the Cosmic Infrared Background radiation
(CIB) photons, the detection of further BL Lac objects is of utmost
importance. Thus the search and the detailed study of new extragalactic
TeV sources is one of the major objectives of the HEGRA experiment.

The paper is organized as follows:
We describe the data samples used in the analysis in Section 2 and the
method used for computing flux upper limits in Section 3. The extinction
by the CIB is discussed in Section 4, and in Section 5 we give the
experimental results. In Section 6 and 7 we focus on two objects: (1)
1ES 2344+51.4 detected during December 1995 by the WHIPPLE Cherenkov
telescope (Catanese {\it et al.} 1997), 
and  BL Lacertae which, in July 1997, showed
a strong flare detected in gamma-ray by EGRET, in the visible
(Bloom {\it et al.} 1997; Madejski {\it et al.} 1999) and 
in X-ray wavelengths. 
A strategy to optimize the search for new extra-galactic 
sources is briefly outlined in Section 8, and the conclusions
are given in Section 9.
\section{Data}
\label{data}
The HEGRA experiment, located on the Canary Island La~Palma at the
Observatorio del Roque de los Muchachos (2200~m a.s.l., 28.75$^\circ$N,
17.89$^\circ$W), consists of several arrays of particle and 
Cherenkov-detectors dedicated to cosmic ray research (Lindner 1997,
Barrio {\it et al.} 1998). 

Before September 1998, the CT-System consisted of four, 
(now it consists of five) telescopes 
with 8.5~m$^2$ mirror area each.
Each telescope is equipped  
with a 271 pixel camera with a pixel size of 0.25$^\circ$, 
covering a field of view of 4.3$^\circ$.
The cameras are read out by 8 bit 120 Mhz Flash-ADC systems.
Details about the CT-system and the performance of the 
stereoscopic air shower observation method are given by 
(Daum {\it et al.} 1997) and (Aharonian {\it et al.} 1999a).

Table \ref{tab_data}  summarizes the observations and redshifts of the 
nine BL~Lac 
objects considered in this work.
The first 7~objects with right ascension $<$ 9~h have been observed
during October and November 1997. The data
have been taken with only 3 telescopes in the system 
because the telescope CT4 was not 
operational from October 16 to November 15 1997.
The last two sources have been observed with the 4-CT system, BL Lacertae in
July/August 1997 and May/June 1998 and  1ES~2344+51.4 in December 1997.

\begin{table}[hhh]
\begin{center}
\begin{tabular}{cccccc}
\hline
~~~~~Name & ~~$z$ & MJD & Hours & $ \langle \theta \rangle$ & $ n.t.$\\
\hline
1ES0145+13.8 & .125 & 50779.9 &  1.1 & 15.6 & 3\\
\hline
1ES0219+42.8 & .444 & 50752.1 &  1.3 & 19.6 & 3 \\
\hline
1ES0229+20.0 & .139 & 50782.9 &  1.3 & 13.8 & 3 \\
\hline
1E0317.0+1834 &.19 & 50782.0 & 1.3 & 12.5 & 3 \\
\hline
& & 50777.1 &  1.0 &  28.6  & 3\\
1ES0414+00.9             &   .287     & 50783.1 & 1.3 &  28.6 & 3\\
             &        & 50787.0 & 0.9 &  28.0 & 3\\
\hline
1ES0647+25.0 & ~~?&   50780.1 & 0.6 &   6.3 & 3  \\
\hline
 & & 50782.2 &  2.0 &  25.8  & 3\\
             &        & 50783.1 & 4.0 &  30.5  & 3\\
2E0829.1+0439             &    .18    & 50784.1 &4.1 &  31.0 & 3 \\
             &        & 50785.1 & 2.4 &  26.7 & 3\\
             &        & 50787.2 & 2.0 &  26.3 & 3\\
\hline
 &   & 50659.1 & 1.4 &  14.7 & 4 \\
 1ES2200+42.0           &   .069    & 50660.1 & 2.4 &  16.3 & 4\\
in 1997          &      & 50662.1 & 1.4 &  22.9 & 4\\
            &       & 50663.1 & 2.1 &  21.6 & 4\\
    (BL~Lacertae)        &       & 50669.0 & 0.9 &  16.8 & 4\\
            &       & 50673.1 & 1.0 &  15.3 & 4\\
\hline
   &       & 50962.2 & 0.6 &  40.8 & 4\\
 1ES2200+42.0   &       & 50963.2 & 1.7 &  37.3 & 4\\
       in 1998       &       & 50964.2 & 1.7 &  36.9 & 4\\
            &       & 50965.2 & 1.4 &  34.4 & 4\\
\hline
 &   & 50804.9 &  2.4 &  34.5 & 4 \\
            &       & 50805.9  &  2.4 &  33.3 & 4 \\
            &       & 50806.9  &  1.7 &  37.5 & 4 \\
1ES2344+51.4            &  .044     & 50807.9  &  2.7 &  36.2  & 4\\
            &       & 50808.9  &  2.0 &  34.2  & 4\\
            &       & 50809.9  &  2.0 & 36.3  & 4\\
            &       & 50810.9  &  2.0 & 35.1 & 4\\
            &       & 50812.8  & 0.6 & 31.4   & 4 \\
\hline
\end{tabular}
\end{center}
\caption{Summary of the data samples for the 9 BL~Lac objects. $\langle \theta
\rangle$ denotes the mean zenith angle of the observations and $n.t.$ 
the number of telescopes in the stereoscopic system.}
\label{tab_data} \end{table}
\section{Upper limits in Crab units and in flux units} \label{method}
 In the following analysis, measured fluxes or flux upper limits are first
determined in Crab units and are only subsequently converted into absolute
flux values or flux upper limits. Results in Crab units have the
advantage of relying exclusively on measured data and of being free
from systematic errors due to the Monte Carlo simulations of the air
showers and the detector response. 
Moreover, as several Bl~Lac objects have been observed with only three 
telescopes in the CT-system, we can directly compare their observations
to Crab data taken under the same conditions, at the same epoch,
 avoiding thus specific simulations. The trigger rate is constant within 5\%
for a given zenith angle (Aharonian {\it et al.} 1999a), thus
no strictly simultaneous data
can be compared safely.

For each source and each zenith angle interval $I_{za}$
the number of events in the ON-source region ($ON_s$), 
the number of events in the OFF-source region ($OFF_s$), and 
the observation time $T_s$ is determined. To maximize
the statistics the analysis is based on
``loose'' cuts: the mean scaled width of the showers 
(Konopelko 1995;  Daum {\it et al.} 1997) has
to be smaller than 1.2 
(to retain $\approx$~80\% of photon induced showers) 
and the squared angular
distance of the reconstructed shower direction from the source 
direction has to be smaller
than 0.05~deg$^2$.  
For each time period of fixed experimental conditions, a reference Crab data
sample was analyzed (compare Table \ref{tab_Crab}) 
and the numbers $ON_c$, $OFF_c$, and
$T_c$ were determined for all zenith angle intervals. 
As Crab observations with three telescopes have been performed only for 
several days, the flux upper limits derived with this reference data 
are considerably higher than for the data with 4 telescopes in the system.

Using the probability density function of the number of source events, 
we compute  
the upper limit of the number of counts $MAX_s$ 
from $T_s$ hours of source observations 
 at 99\%~confidence level (Helene 1983).
Similarly, we calculate the lower limit $MIN_c$ of the  number of counts
for $T_c$ hours of  Crab observations. 
We compute the upper limit in Crab units $UL_s$ from:
$ UL_s = \frac{MAX_s}{MIN_c}  \times \frac{T_c}{T_s}$.

The energy threshold $E_{th}$ is computed by Monte-Carlo
for each source as a function of 
the mean zenith angle of the object during the observations.
For the CT-system, 
the energy threshold scales with zenith angle $\theta$ roughly as
$\cos^{-2.5}(\theta)$ (Konopelko {\it et al.} 1998). 
Only data with good weather conditions are used and phototube 
voltage fluctuations are corrected. 

\begin{table}[hhh] 
\begin{center} 
\begin{tabular}{cc} 
\hline 
MJD of the Crab observations & hours  \\
\hline
\multicolumn{1}{c}{\it 3 telescopes} &\\
\hline
50747 to 50750, 50760, 50777, 50778 & 15h40\\
\hline
\multicolumn{1}{c}{\it 4 telescopes} &\\
\hline
50721, 50731, 50732, 50788, 50789, 50806, 50831&   \\
50809 to 50811, 50835 to 50837, 50864 to 50866&  54h45\\
50869, 50872 to 50874, 50891 to 50893, 50902 & \\
\hline
\end{tabular}
\end{center}
\caption{Information about the observations of the Crab Nebula  
used for computing flux upper limits in Crab units. } 
\label{tab_Crab} \end{table} 

Assuming a source energy spectrum, the conversion of upper limits 
in Crab units into upper limits in absolute flux units is straightforward. 
In the following we use two slopes for observed source spectrum:
dN/dE $\propto \ E^{-2.6}$, as measured for the Crab nebulae around 1~TeV
(Konopelko {\it et al.} 1998) and a steeper one
dN/dE $\propto\ E^{-3.6}$.
Above 0.5 TeV, the integral Crab flux is
$F_{-2.6}$= \mbox{$5 \cdot 10^{-11}$~cm$^{-2}$s$^{-1}$}.
In the other case, we normalise the flux in order to get the same
integral flux in this range.
\\
The upper limits on the integral flux are then computed from:
\\ $UL_{-2.6} = UL_s \times 1.7 \cdot 10^{-11}\ E_{th}^{-1.6}\ {\rm cm}^{-2}\ {\rm s}^{-1}$, and 
\\   $UL_{-3.6} = UL_s \times 8.5 \cdot 10^{-12}\ E_{th}^{-2.6}\ {\rm cm}^{-2}\ {\rm s}^{-1}$.
\section{Correction for intergalactic absorption}
\label{cirb}
\begin{figure}
\includegraphics[width=9cm]{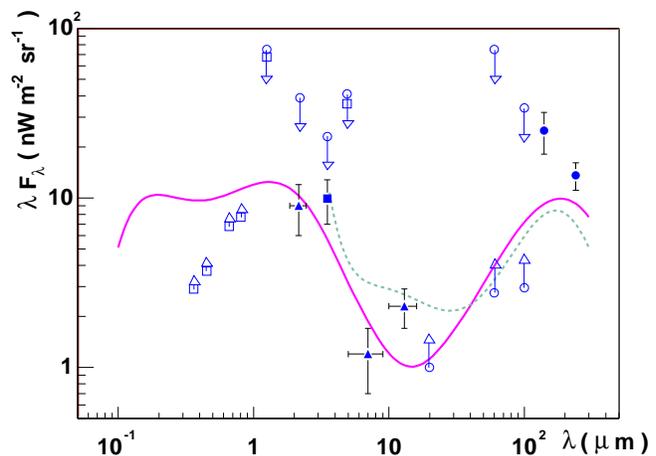}
\caption{
Experimental constraints on the CIB energy density together with
LCDM  model by Primack {\it et al.} 1999 (full line)
 and the maximum CIB given by Malkan \& Stecker 1998 (dashed line). 
The triangles show ISO measurements at mid-infra-red 
(Stanev {\it et al.} 1998); 
the open circle shows a tentative detection 
at 3.5$\mu m$ (Dwek {\it et al.} 1998a); 
all other measurements, lower and upper limits
are from a recent compilation by (Dwek {\it et al.},1998b).
}\label{fig_cirb}
\end{figure}

%
Interaction of TeV gamma-rays with intergalactic infrared photons
by pair production processes probably modifies substantially 
the intrinsic TeV spectrum emitted by the source (Nikishov 1962; Gould \& Schreder 1965;
 Stecker {\it et al.} 1992).
The most important domain for us is 1-10 $\mu$m as these photons can interact
 with TeV gamma-rays around 
the HEGRA threshold. The curve in Fig.~\ref{fig_cirb} presents upper limits and measurements
in relative agreement with recent 
modelling of CIB (Malkan \& Stecker 1998; Primack {\it et al.} 1999).
Gamma photons below 2~TeV effectively interact only with CIB photons below few 
$\mu$m, so only Primack models
can be used.
As the purpose of this section is  to confirm ideas, we choose the  LCDM cosmology model 
(cosmological constant + cold dark matter) which gives higher density and is therefore
more conservative than the HCDM one (hot+cold dark matter).

\begin{figure}
\includegraphics[width=9cm]{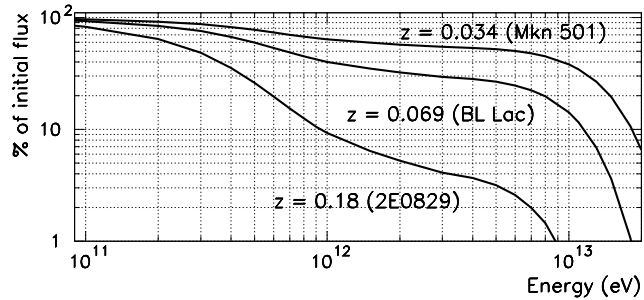}
\caption{Percentage of VHE photons reaching the observer after
absorption by the CIB using LCDM  model by Primack {\it et al.} 1999.}\label{fig_tau}
\end{figure}

For a non-evolving CIB density according to this model,
the amount of absorption for 0.1 to 20~TeV photons is shown in
Fig. \ref{fig_tau} for three of the studied BL~Lac objects.
Strictly speaking, the CIB energy density depends on redshift due to
evolutionary effects, which are not well known. 
However, the redshift dependence could be neglected for redshifts 
lower than 0.15 since we do not expect significant evolution on very 
short time scales. 

Between 500~GeV and 1~TeV, 25\% 
of the flux is absorbed even for a source as close as Mkn~501 with z=0.034; 
for the source 2E0829.1+0439 with a redshift of 0.18, 
this percentage rises to 75\%.
The figure shows that the CIB absorption becomes very large for 
objects with redshifts above $\approx 0.15$.

\section{Experimental results}
\label{limits}
The observations described in this paper do not reveal 
positive evidence for VHE emission from any of the nine studied sources.
Table \ref{tab_lim} summarizes the upper limits in Crab units $UL_s$, 
the mean energy thresholds of the observations, and
the upper limits on the integral fluxes, 
assuming the two different slopes for the spectrum,
 above the energy threshold 
with and without correction of the absorption by the CIB (Primack LCDM model).
The threshold being defined as the product
of the acceptance of the detector and the source gamma-ray  spectrum, it effectively depends on
 the slope of the gamma-ray flux. So the table give thresholds for the Crab slope (-2.6), 
and an effective threshold 30\%
 lower is used for computation of the steeper model. 

All upper limits have a confidence level of
99\%. As mentioned above, the flux upper limits in Crab units 
are free from systematic errors. 
The major systematic uncertainty of the upper limits
in absolute flux units derives from a 15\% uncertainty 
in the energy scale of the CT-System. This 15\% uncertainty
translates into a flux uncertainty of approximately 20\% and 40\%
for a integral source spectral index of -1.6 and -2.6
respectively.
\\
\\
Upper limits of a few times $10^{-12}$~cm$^{-2}$~s$^{-1}$ are 
obtained after a few hours of observation.
The correction of the absorption by the CIB of Fig.\ref{fig_cirb} 
(computed up to 3 TeV)
increases the upper limits by a factor of 2 to 60 (for a slope of -2.6)
depending on the effective threshold and the source redshift. 

Of course
the effect of the CIB absorption  makes the observed spectrum
steeper and then the impact of a very hard intrinsic spectrum
is attenuated (upper limits for the two slopes are less different after absortion
than before).
 We note that the upper limits are derived from assumed
slopes
for spectra and a model for the CIB intensity. These uncertainties make upper limits
rather qualitative.  

\begin{table}[hhh]
\begin{center}
\begin{tabular}{ccccccc}
\hline
(a)  & (b) &(c)  & \multicolumn{2}{c}{$UL_{int}$} (d) & \multicolumn{2}{c}{$UL_{IR1}$} (e) \\
 R.A.  & $E_{th}$  &$UL_s$  & $E^{-2.6}$ &$E^{-3.6}$ &$E^{-2.6}$ & $E^{-3.6}$ \\ 
\hline 
0145 & 580 & .23 & 9.3 & 24.2 &  42.1 & 71.9 \\ 
0219 & 630 & .40 & 14.2 & 30.5 & 6306.3 & 4251.8  \\ 
0229 & 540 & .25 & 11.4 & 35.1 & 58.9 & 106.9\\ 
0317 & 530 & .42 & 19.7 & 63.8 & 195.1 & 296.4 \\ 
0414 & 910 & .65 & 12.8 & 13.5 &   789.7 & 602.0 \\ 
0647 & 510 & .31 & 15.5 & 54.4 &  ? & ? \\ 
0829 & 840 & .14 &  3.1 & 3.8 & 37.1 & 35.6  \\ 
2200a & 580 & .11 &  4.5 & 11.6 &  9.8 & 20.7 \\ 
2200b & 870 & .15 &  3.2 & 3.6 &  7.4 & 8.4\\ 
2344 & 990 & .10 &  1.7 & 1.6 & 2.8 & 2.7 \\ 
\hline
\end{tabular}
\end{center}
\caption{99\% confidence level upper limits 
derived in this work:
(a) right ascension of the source, 
2200a designates BL~Lacertae in 1997, and 2200b BL~Lacertae in 1998, 
(b) mean energy threshold of the observations in GeV for the -2.6 slope
(the effective threshold is approximately 200 GeV lower for the -3.6
slope),
(c) upper limits in Crab units, 
(d) upper limits on the integral flux above the threshold energy
in units of $10^{-12}$~cm$^{-2}$~s$^{-1}$, and 
(e) upper limits on the integral flux between the threshold energy and 3~TeV
after correction for the CIB absorption (using  LCDM model by Primack {\it et al} 1999)
in units of $10^{-12}$~cm$^{-2}$~s$^{-1}$.
The slope of the assumed differential 
spectrum is -2.6 or -3.6 (see text for details).}
\label{tab_lim}
\end{table}
\section{BL~Lacertae (1ES 2200+42.0)} \label{bllac}
\begin{figure}
\includegraphics[width=9cm]{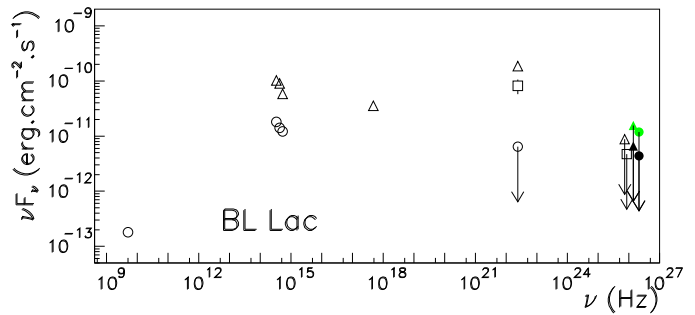}
\caption{Spectral energy distribution of the object BL~Lacertae. 
Data are taken from (Perlman {\it et al.} 1996;
Webb {\it et al.} 1998; Catanese {\it et al.} 1997); 
different measurements well before and after the 1995 and 
the 1997 flare (circles), measurements around the 1995 flare (squares) 
and  measurements around the 1997 flare (triangles).
HEGRA CT-system data are represented by filled symbols while
others are represented by open ones.
The black symbols indicate the raw limit, the grey ones the limit after deconvolution
by the CIB absorption (LCDM  model by Primack {\it et al.} 1999).
}\label{fig_sed_bllac}
\end{figure}

Several experiments extensively observed this source following the detection 
of a strong flare by EGRET in the $\gamma$-ray regime (only several hours
observation time) and simultaneously in the optical regime 
around July 19th, 1997 (MJD 50648)
(Bloom {\it et al.} 1997; Madejski {\it et al.} 1999). 
Because this flare occurred during a full moon period, the TeV observations 
started only about ten days after the detection of the flare. 
Our observations yielded a flux upper limit
of $4.5 \cdot 10^{-12}$~ph~cm$^{-2}$~s$^{-1}$ ($E>$580~GeV).
The CAT group obtained a similar upper limit:
$11.4 \cdot 10^{-12}$~ph~cm$^{-2}$~s$^{-1}$ 
($\approx$11\% of the Crab) ($E>$300~GeV)
(Barrau, private communication). 
Earlier measurements at VHE energies were performed in 1995 with
the Whipple telescope following a weak flare detected by EGRET with
an integral $\gamma$-ray flux of 
$4\pm1.2 \cdot 10^{-7}$~ph~cm$^{-2}$~s$^{-1} $ ($E>$100~MeV), 
i.e. three times lower than the flux of the July, 1997 flare. 
The Whipple observations (40 h) yielded a flux upper limit of
5.3~10$^{-12}$~ph~cm$^{-2}$~s$^{-1}$ ($E>$350~GeV) 
(Catanese {\it et al.} 1997). 
The results from the three VHE experiments HEGRA, CAT, and Whipple are
shown in Fig. \ref{fig_sed_bllac}.

Information about the
X-ray activity between February 1996 and August 1998 in the energy region  2-12 keV is provided
by the All Sky Monitor (ASM) on board the {\it Rossi X-Ray Timing
Explorer} (RXTE) (Remillard 1997).  
We determined the ASM count rates from the 
``definitive'' results obtained through
analysis of the processed data by the RXTE ASM team at MIT;
data have a dwell duration larger than 30~seconds 
and a flux fit with a reduced $\chi^2$-value below 1.5. 
The light curve is extracted using the ``ftools 4.0'' package.
Unfortunately, the mean count rates are very low, {\it i.e.} typically
around $10^{-2}$-$10^{-1}$~Hz and the observation frequency is less
1 ``run'' of 90 seconds per hour, 
making it difficult to get information about
the source activity on time scales as short as one day.
So data are binned in 5 week bins, yielding
flux estimates with acceptable statistical errors.

\begin{figure}
\includegraphics[height=9cm,angle=-90]{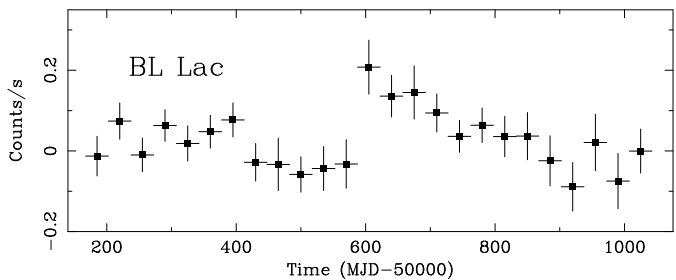}
\caption{Light curves computed from data of the 
ASM/RXTE detector (1.3-12.1~keV) with a binning of five weeks. 
Observations range from February 1996 to August 1998.
}\label{fig_asm_bl}
\end{figure}

The light curve is shown in Fig. \ref{fig_asm_bl}. 
A pronounced luminosity increase during
June 1997 followed by a slow decrease during the following 5 months
can be recognized. HEGRA data were taken in July/August 1997 when the source was still active.
In 1998, Bl~Lacertae was in a low X-ray activity. 

Note that EGRET was able to detect the source, 
and the TeV instruments were not. 
Intriguingly in the case of Mkn 501 the opposite happened:
the source was bright in the TeV energy regime but could hardly be
detected at MeV/GeV energies.

Model calculations by B\"ottcher \& Bloom (1998) and Madejski (1999) predict  TeV flux
significantly higher than the upper limits obtained by HEGRA 
(and also by CAT at the same period).
We explain the non-detection at VHE energies by two 
reasons. First, the absorption by a CIB density as in Fig.\ \ref{fig_cirb}
reduces the observable flux by about 50\%. 
Secondly, the HEGRA measurements were taken roughly 10 days 
after the detection of the GeV-flare. 
Taking into account the strong variability observed in the two 
BL Lac objects Mkn 501 and Mkn 421, 
the VHE flux of BL Lacertae could have easily decreased by a factor of 
$>$5 in 10 days.

\section{1ES 2344+51.4} \label{2344}
\begin{figure}
\includegraphics[width=9cm]{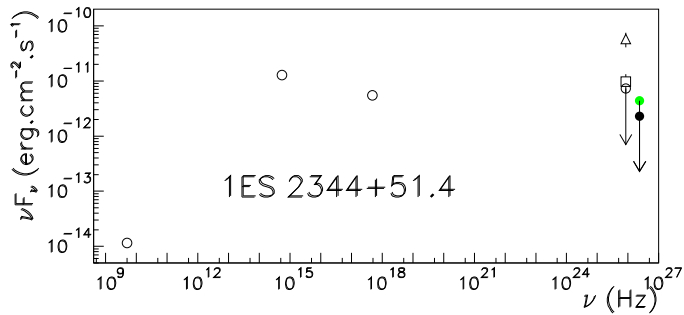}
\caption{Spectral energy distribution of the source 1ES~2344+51.4. 
Data are taken from
Catanese {\it et al.} 1998 and Qin and Xie 1997: 
measurements of different periods before and after the
1995/96 winter period (circles), 
mean flux value of the high state during the 1995/1996 winter 
period (square) and flux during the 1 day flare observed by 
Whipple (triangle). 
The HEGRA CT-system upper limit is shown as a filled symbol 
while complementary data are represented by open symbols.
The black dot indicates the raw limit, the grey one the limit after deconvolution
by the CIB absorption (LCDM  model by Primack {\it et al.} 1999).
}\label{fig_sed_2344}
\end{figure}

The spectral energy distribution of 1ES~2344+51.4 is shown 
in Fig. \ref{fig_sed_2344}. This source was observed by
the Whipple group during the winter months of 1995/96 and 1996/97
(Catanese {\it et al.} 1998). 
Whipple detected a flare on December 20$^{th}$ 1995 with a flux of
6.6$\pm1.9\ 10^{-11}$~ph~cm$^{-2}$~s$^{-1}$ ($\approx$ 63\% of
the Crab); the mean flux (excluding the flare emission) 
during the first winter was estimated as
1.1$\pm0.4\ 10^{-11}$~ph~cm$^{-2}$~s$^{-1}$ 
($\approx$11\% of the Crab). For the period 96-97
an upper limit of 8.2$\ 10^{-12}$~ph~cm$^{-2}$~s$^{-1}$ 
($\approx$8\% of the Crab) was reported 
(all integral fluxes above 350~GeV).

\begin{figure}
\includegraphics[height=9cm,angle=-90]{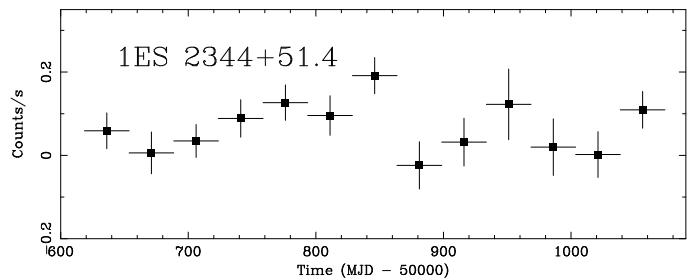}
\caption{Light curves computed from data of the 
ASM/RXTE detector (1.3-12.1~keV) with a binning of five weeks. 
Observations range from June 1997 to August 1998.
}\label{fig_asm_2344}
\end{figure}

The flux limit computed from the 
1997/1998 HEGRA CT-system data is 
2.9$\  10^{-12}$~ph~cm$^{-2}$~s$^{-1}$ ($E>$ 1~TeV). 
The smooth curve in the 2-10~keV band (Fig. \ref{fig_asm_2344}) indicates
that the source has been in a low and stable state for 1.5~years. 
The source was not monitored by the ASM during 
the winter 1995/1996, therefore, unfortunately, the  2-10 keV activity during the state 
of VHE-emission could not be examined.

\section{Observational strategy} \label{strategy}
It is widely believed, that TeV gamma-rays of BL~Lac objects are produced 
by ultra-relativistic electrons, which emit synchrotron radiation in the keV band
and produce the TeV photons due to Inverse Compton scattering of soft
target photons ({\it e.g.}, Ghisellini {\it et al.} 1998).
The X-ray activity of a source gives crucial information about the presence of
high energy electrons which could also produce TeV radiation.
A major uncertainty arises since the keV photons could either be synchrotron
photons of very high energy electrons 
(the actual energy of the electrons certainly depends on the value of the 
magnetic field) or could be inverse Compton photons produced by electrons 
with moderate energies. In the latter case, TeV emission
of the source would be  less probable.

The ASM/RXTE data provide a unique tool for monitoring the light curve
of many X-BL~Lac objects, but with a moderate sensitivity.
In the case of Mkn 501, the state of increased activity lasted 
several months (Aharonian {\it et al.} 1999a).
The ASM data could be used to detect such a state 
by averaging the data over several days or weeks.
Such a large bining time does not allow us to search for short-term variability
but is dictated by the low sensitivity of the detector.
Moreover, as for Mkn~421, the range 2-12~keV
can be located around the ``pivot point'' of the synchrotron spectrum (the hardening
 starts at the energy of the pivot point): it is not a
completely safe indicator of activity, but it is certainly the best one available.
Whenever such a state was detected, observations could be initiated.
Such a strategy could be complemented with an alert-system
which allows us to react to a flare detection in any chosen wavelength
within several hours.

\begin{figure} 
\includegraphics[width=9cm]{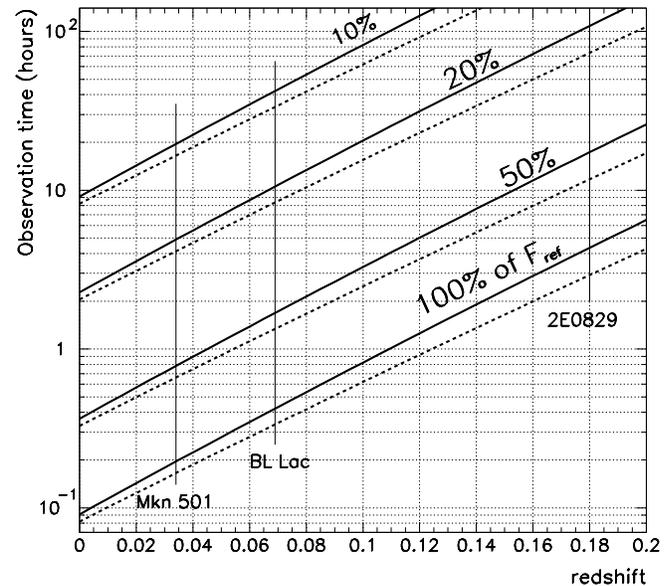} 
\caption{Expected observation time for achieving a 3-$\sigma$ detection 
for a flux (before absorption by the CIB) of 10\%, 20\%, 50\% or 100\%
of $F_{-2.6}$ (full line) or of $F_{-3.6}$ (dashed line). 
Features of the HEGRA 4 CT-system and an energy range of 0.5-3~TeV have
 been assumed for the calculations;
the results are valid for observations with zenith angles below 
$\approx$20$^\circ$.}\label{fig_t_obs} \end{figure}

Fig. \ref{fig_t_obs} shows the observation
time required to achieve a detection for sources with 
redshifts ranging from 0. to 0.2.
The intrinsic emission is assumed to be 
10\% to 100\% of $F_{-2.6}$ (assuming a slope of 2.6) or of $F_{-3.6}$ 
(with a slope of 3.6, see Section 3).
It can be recognized that, depending on the redshift of the source,
observation times of several hours could suffice for a significant detection.
\section{Conclusions}
In this paper we report upper limits for a sample of eight X-ray selected 
BL~Lac objects and the object ``BL Lacertae'' 
observed in 1997/1998 with the HEGRA CT-system.
Seven sources were studied with three telescopes during one to ten
hours.  The TeV upper limits range from 15\% to 65\% 
of the Crab flux. The conversion in units of cm$^{-2}$ s$^{-1}$ is done assuming 
the slope of the spectrum and a model of CIB absorption.

The  sources BL~Lacertae and 1ES~2344+51.4 were observed 
with the 4-CT-system for approximately 15 h each. 
BL Lacertae was flaring in July, 1997 in optical and soft gamma-ray. The source 1ES~2344+51.4 is a very good
VHE candidate due to its similarity  to the 
well established strong TeV sources Mkn~421 and Mkn~501. 
Our upper limits are at the level of $\approx$10\% of the Crab flux.

The HEGRA IACT system could detect TeV sources out
to a redshift of 0.15. Beyond this, intergalactic absorption
is expected to reduce the $>$500~GeV flux considerably.
Since BL Lac sources are known to be very variable, a search for TeV emission
is much more promising when guided by observations in other wavelengths,
especially by observations in the X-ray energy band.
\begin{acknowledgements}
The support of the German Ministry for Research and
Technology BMBF and of the Spanish Research Council CICYT is gratefully
acknowledged.  We thank the Instituto de Astrofisica de Canarias (IAC) for
supplying excellent working conditions at La~Palma. The ASM/RXTE results
are provided by the ASM/RXTE teams at MIT and at the RXTE SOF and GOF at
NASA's GSFC. We acknowledge Ron Remillard and Meg Urry for their helpful
discussions and advice. We also thank  the referee for his helpful remarks.
\end{acknowledgements}

\end{document}